\documentclass[preprint,prstab,showpacs,preprintnumbers,amsmath,amssymb]{revtex4}
\usepackage{epsf}
\usepackage{dcolumn}
\usepackage{bm}
\everymath{\displaystyle}
\begin{document}

\title{Equation of spin motion in storage rings in a cylindrical coordinate system}

\author{Alexander J. Silenko}
\affiliation{Institute of Nuclear Problems, Belarusian State
University, 220080 Minsk, Belarus}
\date{\today}

\begin {abstract}
The exact equation of spin motion in a cylindrical coordinate
system with allowance for electric dipole moments of particles has
been derived. This equation is convenient for analytical
calculations of spin dynamics in circular storage rings when the
configuration of main fields is simple enough. The generalized
formula for the influence of a vertical betatron oscillation on
the angular velocity of spin rotation has been found. This formula
agrees with the previously obtained result and contains an
additional oscillatory term that can be used for fitting. The
relative importance of terms in the equation of spin motion is
discussed.

\end{abstract}

\maketitle

\section {Introduction}

New experiments for measurement of the muon anomalous magnetic
moment (AMM) \cite{gmt,FS} and the electric dipole moments (EDMs)
of fundamental particles \cite{EDM,EDMP} in storage rings need an
extremely high accuracy of measurement of the spin precession in
storage rings. To distinguish the spin precession induced by the
EDMs of particles and other minor effects, the spin motion has to
be mathematically described in a convenient manner. The goal of
this paper is to derive the equation in cylindrical coordinates
which allows an exact analytical description of spin dynamics in
storage rings.
We
do not confine ourselves to specific configurations of electric
and magnetic fields. In particular, electric or magnetic focusing
fields and an accelerating electric field can be used. The
influence of EDMs on the spin dynamics is taken into account. We
consider the problem of averaging the angular velocity of spin
rotation and calculate the generalized formula for the influence
of vertical betatron oscillation upon this quantity.

There are many algorithms of computer calculations based on
different coordinate systems (e.g., on Frenet-Serret coordinates
\cite{CR}). These algorithms cover any problem of beam and spin
dynamics in storage rings. However, an analytical solution of the
problem can be necessary for several high-precision experiments.
We mean the g$-$2 \cite{gmt,FS} and EDM \cite{EDM,EDMP}
experiments, those precision is extremely high. These experiments
need a clear understanding of spin dynamics. The use of
cylindrical coordinates for analytical calculations of spin
dynamics can be very successful if the configuration of main
fields is simple enough. For circular storage rings, it is quite
natural to describe the spin motion just in cylindrical
coordinates. For this reason, other formalisms are less convenient
for the g$-$2 and EDM experiments. This problem will be considered
properly later.

The relativistic system of units $\hbar=c=1$ is used.

\section {GENERAL EQUATIONS OF PARTICLE AND SPIN MOTION}

The particle trajectory in a circular storage ring is
approximately a circle. Coherent betatron oscillations (CBOs) of a
beam as a whole in the horizontal and vertical planes affect the
trajectories of individual particles. Synchrotron motion occurs
whenever radio frequency (rf) cavities are used. The incoherent
motion of particles can also take place. As a result, the particle
trajectories are never closed. Field defects and misalignments of
magnets, electric field plates and electrostatic quadrupoles cause
particle trajectory distortions. These effects lead to a change in
the magnetic field acting on the spin in the particle rest frame.
As a result, the spin motion becomes complicated. For this reason,
equations of spin motion correctly taking into account
perturbations of the particle trajectory have only been obtained
for some specific problems \cite{GF,FPL,FF}. In Refs.
\cite{FPL,FF}, the corrections to the spin rotation frequency due
to the vertical and radial betatron oscillations have been
calculated.

As a rule, the use of the one-particle approximation is sufficient
to describe the particle and spin motion. In this approximation,
coherent and incoherent betatron oscillations give similar
effects. However, the CBOs are more important than the incoherent
BOs because they cause a systematic shift in the spin precession
frequency in the real experiments (see Refs. \cite{gmt,FS}).

The particle motion is described by the Lorentz equation
\begin{equation}
\frac{d\bm p}{dt}=e\left(\bm E+\bm\beta\times\bm B\right), ~~~
\bm\beta=\frac{\bm v}{c}=\frac{\bm p}{\gamma m}.
\label{eq1}\end{equation}

It is convenient to use the unit vector of momentum direction $\bm
N=\bm p/p$ which defines the direction of particle motion. Since
$$\frac{d\bm N}{dt}=\frac{\dot{\bm p}}{p}
-\frac{\bm p}{p^3}\left(\bm p\cdot\dot{\bm p}\right),$$ Eq.
(\ref{eq1}) takes the form \begin{equation}\frac{d\bm
N}{dt}=\bm\omega\times\bm N, ~~~ \bm\omega=-\frac{e}{\gamma
m}\left(\bm B-\frac{\bm N\times\bm E}{\beta}\right),
\label{eqm}\end{equation} where $\bm\omega$ is the angular
velocity of particle rotation.

The spin motion is determined by the
Thomas-Bargmann-Michel-Telegdi (T-BMT) equation
\begin{equation}\begin{array}{c}
\frac{d\bm s}{dt}=\bm\Omega_{T-BMT}\times\bm s,
 \\
\bm\Omega_{T-BMT}=-\frac{e}{2m}
\left\{\left(g-2+\frac{2}{\gamma}\right)\bm B-
\frac{(g-2)\gamma}{\gamma+1}\bm\beta(\bm \beta\cdot \bm B)
-\left(g-2+\frac{2}{\gamma+1}\right)(\bm\beta\times\bm E)\right\},
\end{array}\label{T-BMT}\end{equation} where $\bm s$ is the spin vector.
This equation was derived by Thomas \cite{T} (also by Frenkel
\cite{F}) and, in a more general form, by Bargmann, Michel and
Telegdi \cite{BMT}. In this work, ``spin" means an expectation
value of a quantum mechanical spin operator.

The T-BMT equation describes the motion of spin in the rest frame
of the particle, wherein the spin is described by a
three-component vector. The longitudinal direction in this frame
is mapped from the longitudinal direction in the laboratory frame.

A comparison of Eqs. (\ref{eqm}) and (\ref{T-BMT}) shows that the
spin of the particle placed in a vertical magnetic field rotates
in the horizontal plane with a frequency proportional to g$-$2. We
use the term ``g$-$2 precession" for any spin rotation in the
horizontal plane, even in the presence of an electric field.

\section {CORRECTIONS TO THE ANGULAR VELOCITY OF
PARTICLE MOTION IN THE HORIZONTAL PLANE}

The particle spin motion in storage rings is usually specified
with respect to the particle trajectory. 
The use of the cylindrical coordinates considerably simplifies the
analysis of g$-$2 precession and effects caused by the particle
EDM. The particle rest frame is less convenient for taking into
account the acceleration of the particle and the tilt of its
orbit. When the configuration of the main fields is simple enough,
the use of cylindrical coordinates simplifies the form of
equations of particle and spin motion. However, the description of
the spin motion in the cylindrical coordinate system is a
difficult problem because the axes of this system are defined by
the position of the particle being in oscillatory motion. The
transformation of the T-BMT equation to the cylindrical
coordinates should be performed with allowance for the oscillatory
terms in the particle motion equation.

The vertical CBO and orbit distortions change the plane of
particle motion. The (pseudo)vector of angular velocity becomes
tilted. The instantaneous plane of particle motion does not
coincide with the horizontal plane, and the instantaneous plane of
rotation of vector $\bm N$ is not horizontal. The angle $\Phi$
between two positions of the rotating vector $\bm N$ in the tilted
plane is not equal to the angle $\phi$ between two corresponding
horizontal projections. Therefore, the vertical CBO and the orbit
distortions change the instantaneous angular velocity of particle
motion. This effect can be calculated.

We suppose the $x$- and $y$-axes are horizontal and the $z$-axis
is vertical. Since we use the cylindrical coordinate system, it is
convenient to direct the $z$-axis orthogonally to the plane of the
unperturbed particle motion. We can define the angle of particle
rotation in the $xy$-plane as the angle between two horizontal
projections of the vector of momentum direction in two indicated
positions, $\bm N_\|$ and $\bm N'_\|$. The infinitesimal angle of
particle rotation in the $xy$-plane, $d \phi$, is given by
$$d\phi=\frac{(\bm N_\|\times\bm N'_\|)\cdot\bm
e_z}{|\bm N_\| |\cdot|\bm N'_\| |}=\frac{(\bm N_\|\times d\bm
N_\|)\cdot\bm e_z}{|\bm N_\| |^2}, $$ where $d\bm N_\|=\bm
N'_\|-\bm N_\|$ and $d\bm N_\|$ is an infinitesimal vector. The
sign $\|$ means a horizontal projection for any vector. The
quantity $d\bm N_\|$ determines the deflection of the particle
momentum direction in a time interval $dt$. The
instantaneous angular velocity of particle 
rotation in the horizontal plane is equal to
\begin{equation}
\dot{\phi}\equiv\frac{d\phi}{dt}=\frac{(\bm N_\|\times\dot{\bm
N}_\|)\cdot\bm e_z}{|\bm N_\| |^2}=\omega_z-o,
\label{eq2}\end{equation} where
\begin{equation}
o=\frac{(\omega_xN_x+\omega_yN_y)N_z}{1-N_z^2}=\frac{(\omega_\rho
N_\rho+\omega_\phi N_\phi)N_z}{1-N_z^2}. \label{eqo}\end{equation}
Different components of vector $\bm\omega$ are defined by Eq.
(\ref{eqm}). Indexes $\rho$ and $\phi$ mean projections onto the
basis vectors $\bm e_\rho$ and $\bm e_\phi$ of the cylindrical
coordinate system.

Eqs. (\ref{eq2}),(\ref{eqo}) are exact. We can confirm the
validity of these equations by the fact that they perfectly
describe a constant tilt of the particle orbit and then evaluate
the quantity $o$ in experimental conditions when the tilt of the
particle orbit oscillates.

If the particle orbit is perfectly horizontal, the momentum
direction is defined by
$$\begin{array}{c}
N_x=-\sin{(\omega_0t+\varphi_0)}, ~~~
N_y=\cos{(\omega_0t+\varphi_0)}, ~~~
N_z=0,\end{array}$$ where $\omega_0$ is the cyclotron frequency
and $\varphi_0$ is an arbitrary phase. If the normal to the tilted
particle orbit is orthogonal to the $y$-axis and deflected from
the $z$-axis by a constant angle $\theta$, $y$-components of all
vectors remain unchanged and the components of vectors $\bm\omega$
and $\bm N$ are equal to
\begin{equation}\begin{array}{c}
\omega_x=\omega_0\sin{\theta}, ~~~
\omega_y=0, ~~~
\omega_z=\omega_0\cos{\theta},\\
N_x=-\cos{\theta}\sin{(\omega_0t+\varphi_0)}, ~~~
N_y=\cos{(\omega_0t+\varphi_0)}, ~~~
N_z=\sin{\theta}\sin{(\omega_0t+\varphi_0)}.\end{array}\label{eqdef}\end{equation}
In this case, Eq. (\ref{eq2}) takes the form
\begin{equation}
\dot{\phi}=\frac{\omega_0\cos{\theta}}{1-\sin^2{\theta}\sin^2{(\omega_0t+\varphi_0)}}.
\label{eqt}\end{equation}

As a result of integrating and averaging over time, we obtain
$<\dot{\phi}>=\omega_0$. This result agrees with the evident fact
that the average frequencies of particle motion in the tilted and
horizontal planes are equal and therefore confirms the validity of
Eqs. (\ref{eq2}),(\ref{eqo}).

To evaluate the quantity $o$ in real experimental conditions, we
may confine ourselves to taking into account only the particle
trajectory perturbations caused by the radial and vertical CBOs.
The synchrotron motion does not affect this quantity. By our
estimate, it is sufficient to approximate the horizontal and
vertical betatron motion by the simple forms
$$\begin{array}{c}
N_\rho=\frac{p_\rho}{p}=\rho_0\sin{(\omega_rt+\alpha)},\\
N_z=\frac{p_z}{p}=\psi_0\sin{(\omega_vt+\delta)},
\end{array}$$ where $\rho_0$ and $\psi_0$ are angular amplitudes, $\omega_r$ and $\omega_v$ are
angular frequencies of the radial and vertical CBOs, respectively.

With allowance for orders of quantities
\begin{equation}\begin{array}{c}\omega_\rho\sim\psi_0\omega_v, ~~ \omega_\phi\sim \rho_0\psi_0\omega_v, ~~
N_\rho\sim\rho_0, ~~ N_\phi\approx\pm1, ~~ N_z\sim\psi_0,
\end{array}\label{eqmm}\end{equation}
we obtain that the quantity $o$ is of the third order in the
angular amplitudes $\rho_0$ and $\psi_0$. Moreover, it oscillates
and therefore averages to zero. If we take into account only the
second-order terms in the angular amplitudes and the average
particle orbit is not tilted, the quantity $o$ is negligible.
Approximately,
\begin{equation}
\dot{\phi}=\omega_z=-\frac{e}{\gamma m}\left(B_z-\frac{(\bm
N\times\bm E)_z}{\beta}\right). \label{eq3}\end{equation}

\section {EQUATION OF SPIN MOTION IN CYLINDRICAL COORDINATES}

To transform the general equation of spin motion (\ref{T-BMT}) to
cylindrical coordinates, we need to find the quantities
$$\frac{ds_\rho}{dt}, ~~~ \frac{ds_\phi}{dt}, ~~~ {\rm and} ~~~
\frac{ds_z}{dt}.$$ It follows from the geometry of the problem
that the horizontal axes $\bm e_\rho$ and $\bm e_\phi$ rotate with
the instantaneous angular velocity
$$\bm\omega'=\dot{\phi}\bm e_z. $$


It is easy to show by simple transformations that the spin motion
with respect to the axes of the cylindrical coordinate system can
be written in the form
\begin{equation}
 \frac{d\bm s}{dt}=\bm\omega_a\times\bm s,~~~\bm\omega_a=\bm\Omega-\dot{\phi}\bm
 e_z.
\label{eq5}\end{equation} In this equation, $\bm\omega_a$ is the
the angular velocity of spin rotation in cylindrical coordinates.
The corresponding angular velocity in Cartesian coordinates equals
$\bm\Omega$. If the spin motion is described by the T-BMT
equation, $\bm\Omega=\bm\Omega_{T-BMT}$. The difference between
the quantities $\bm\omega_a$ and $\bm\Omega$ is caused by the
rotation of the axes $\bm e_\rho$ and $\bm e_\phi$.

The equation of spin motion with allowance for the EDM can be
obtained by modifying the T-BMT equation. The EDM and AMM
determine the real and imaginary parts of the same interaction
Lagrangian \cite{FMS,GTh}. This Lagrangian is used to derive an
extended Dirac equation. The contributions of the EDM and AMM to
the Lagrangian are equal to \cite{FMS}
\begin{equation} {\cal L}_{AMM}=\frac{\mu'}{2}\sigma^{\mu\nu}F_{\mu\nu},
~~~~~ {\cal
L}_{EDM}=-i\frac{d}{2}\sigma^{\mu\nu}\gamma^5F_{\mu\nu}, ~~~
 \sigma^{\mu\nu}=\frac{i}{2}\left(\gamma^\mu\gamma^\nu-\gamma^\nu\gamma^\mu\right),
\label{eql} \end{equation} where $\mu'=e(g-2)/(4m)$ is the AMM,
$d$ is the EDM, $\gamma^\mu ~(\mu=0,1,2,3)$ are the Dirac
matrices, and $F_{\mu\nu}$ is the electromagnetic field tensor.
The use of the explicit forms of ${\cal L}_{AMM}$ and ${\cal
L}_{EDM}$ shows that the contribution of the EDM to the full
Lagrangian can be obtained from the corresponding contribution of
the AMM by the substitution \begin{equation} \bm B \rightarrow \bm
E,~~~ \bm E\rightarrow -\bm B,~~~ \mu'\rightarrow d. \label{eqtm}
\end{equation}

Taking into account the particle EDM leads to the following
modification of the T-BMT equation \cite{Sem,NPR,RPJ}:
\begin{eqnarray}
\frac{d\bm s}{dt}=\bm\Omega\times\bm s, ~~~ \bm\Omega=\bm\Omega_{T-BMT}+\bm\Omega_{EDM},
\nonumber\\
\bm\Omega_{EDM}=-\frac{e\eta}{2m}\left(\bm
E-\frac{\gamma}{\gamma+1}\bm\beta(\bm\beta\cdot\bm
E)+\bm\beta\times\bm B\right),\label{eq}\end{eqnarray} where
$\bm\Omega_{T-BMT}$ is defined by Eq.
(\ref{T-BMT}) 
and $\eta=4dm/e$.

The EDM affects the particle motion if the electric field is not
uniform. However, the correction to the equation of particle
motion is negligible. 
If rf cavities are not used, we can also neglect the term
$\frac{\gamma}{\gamma+1}\bm\beta(\bm\beta\cdot\bm E)$.

The transformation of Eq. (\ref{eq}) to cylindrical coordinates is
given by Eq. (\ref{eq5}), where
\begin{eqnarray}
\bm\omega_a=-\frac{e}{m}\left\{a\bm B-
\frac{a\gamma}{\gamma+1}\bm\beta(\bm\beta\cdot\bm B)\right.\nonumber\\
+\left(\frac{1}{\gamma^2-1}-a\right)\left(\bm\beta\times\bm
E\right)+\frac{1}{\gamma}\left[\bm B_\|
-\frac{1}{\beta^2}\left(\bm\beta\times\bm
E\right)_\|\right]\nonumber\\ \left.+ \frac{\eta}{2}\left(\bm
E-\frac{\gamma}{\gamma+1}\bm\beta(\bm\beta\cdot\bm
E)+\bm\beta\!\times\!\bm B\right)\!\right\} +o\bm e_z, ~~~
a=\frac{g-2}{2}. \label{eq7}\end{eqnarray} This formula is exact,
and $\bm\omega_a$ is the angular velocity of spin precession. Eqs.
(\ref{eq5}),(\ref{eq7}) describe the spin motion in arbitrary
storage rings with allowance for the EDM. The term $o\bm e_z$ in
formula (\ref{eq7}) is negligible for the EDM and g$-$2
experiments. Measurements of the AMM in the
g$-$2 experiment are performed at an energy such that 
$1/(\gamma^2-1)= a$, that is $\gamma= 29.3$. In this case, the
third term in Eq. (\ref{eq7}) is equal to zero \cite{gmt,FS}.

After neglecting the small terms, the equation for the angular
velocity of the g$-$2 precession with allowance for the EDM takes
the form
\begin{eqnarray}
\bm\omega_a=-\frac{e}{m}\Biggl\{a\bm B-
\frac{a\gamma}{\gamma+1}\bm\beta(\bm\beta\cdot\bm B)
\nonumber\\
+\left(\frac{1}{\gamma^2-1}-a\right)\left(\bm\beta\times\bm
E\right)\nonumber\\+\frac{1}{\gamma}\left[\bm
B_\|-\frac{1}{\beta^2}\left(\bm\beta\times\bm
E\right)_\|\right]
+ \frac{\eta}{2}\left(\bm E+\bm\beta\times\bm B\right)\Biggr\}.
\label{eq8}\end{eqnarray}

Formulae (\ref{eq7}),(\ref{eq8}) are useful for analytical
calculations of spin dynamics in cylindrical coordinates with
allowance for field misalignments and beam oscillations.

\section {AVERAGING OF THE ANGULAR VELOCITY OF SPIN ROTATION}

The 
spin dynamics given by formula (\ref{eq7}) is
simple when the tilt of the spin out of the $xy$-plane accumulates
from turn to turn. Such a behavior of the spin takes place in the
planned EDM experiment. However, in the g$-$2 experiment one needs
to determine the average and integral characteristics of the spin
motion in the horizontal plane \cite{gmt,FS}. The instantaneous
angular velocity of spin rotation in the horizontal plane,
$\dot{\psi}$, is characterized by the change of angle $\psi$
determining the spin orientation in this plane. The quantity
$\dot{\psi}$ can be found similarly to the corresponding quantity
$\dot{\phi}$ describing the particle rotation (Section III). It is
given by
\begin{equation}
\dot{\psi}\equiv\frac{d\psi}{dt}=\frac{(\bm\xi_\|\times\dot{\bm
\xi}_\|)\cdot\bm e_z}{|\bm\xi_\| |^2}=\left(\omega_a\right)_z-O,
\label{eq1p}\end{equation} where
\begin{equation}
O=\frac{\left[\left(\omega_a\right)_x\xi_x+\left(\omega_a\right)_y\xi_y\right]\xi_z}
{1-\xi_z^2}=\frac{\left[\left(\omega_a\right)_\rho
\xi_\rho+\left(\omega_a\right)_\phi
\xi_\phi\right]\xi_z}{1-\xi_z^2} \label{eq2p}\end{equation} and
$\bm \xi=\bm s/s$.

As an example, we can derive the generalized formula for the
average value of $\dot{\psi}$ affected by the vertical CBO. When
the particle and spin motion is unperturbed, the angular velocity
of spin rotation is equal to
\begin{equation}\Omega_0=\frac {|ea|}{m}B_z.
\label{eqnul}\end{equation} We suppose that $B_z>0$.

The vertical CBO gives an important correction to the g$-$2
frequency (the pitch correction). The formula used currently for
this correction has been calculated by Farley \cite{FPL}. The
result has been confirmed by Field and Fiorentini \cite{FF} and by
computer simulations \cite{YS}. If we ignore the possible
existence of the EDM, then Eq. (\ref{eq8}) can be written in the
form
\begin{equation}
\begin{array}{c} \frac{d\bm
\xi}{dt}=\left\{a_0+a_3\cos{[2(\omega_vt+\delta)]}\right\}(\bm
e_z\times\bm \xi)\\+
 a_2\sin{(\omega_vt+\delta)}(\bm e_\phi\times\bm \xi)
 +a_1\cos{(\omega_vt+\delta)}
 (\bm e_\rho\times\bm
\xi), \end{array}\label{eq9}\end{equation} where
$$\begin{array}{c}
a_0=\lambda\Omega_0\left(1-\frac{\gamma-1}{2\gamma}\psi_0^2\right),\\
a_1=\lambda f\omega_v\psi_0,~~~ a_2=-\Omega_0\frac{\gamma-1}{\gamma}\psi_0,\\
a_3=\lambda\Omega_0\frac{\gamma-1}{2\gamma}\psi_0^2,
\end{array}$$ and $\lambda$ is equal to 1 and $-1$ for
negative and positive muons, respectively. The factor $f$ is given
by
$$
f=1+a\gamma-\frac{1+a}{\gamma}=1+a\beta^2\gamma-\frac{1}{\gamma}
$$ and
$$ f=1+a\gamma $$ for electric and magnetic focusing,
respectively. In the muon g$-$2 experiment, $\psi_0\sim10^{-3}$.
It can be shown that the effect of $a_3$ on the spin dynamics
described by Eq. (\ref{eq9}) is negligible.

If we substitute the zero-order approximation
\begin{equation}\begin{array}{c} \xi_\rho^{(0)}=\xi_\|\cos{(a_0t+\phi_0)}, ~~~
\xi_\phi^{(0)}=\xi_\|\sin{(a_0t+\phi_0)},
~~~
\xi_z^{(0)}=\xi_\bot
\end{array}\label{eqno}\end{equation}
into the right side of Eq. (\ref{eq9}), the first-order
approximation is given by
\begin{equation}
\begin{array}{c} \xi_\rho=\xi_\|\cos{(a_0t+\phi_0)}
+\frac{a_0a_1+
\omega_va_2}{a_0^2-\omega_v^2}\xi_\bot\cos{(\omega_vt+\delta)},\\
\xi_\phi=\xi_\|\sin{(a_0t+\phi_0)}
+\frac{a_0a_2+
\omega_va_1}{a_0^2-\omega_v^2}\xi_\bot\sin{(\omega_vt+\delta)},\\
\xi_z=\xi_\bot
-\left[\frac{a_0a_2+
\omega_va_1}{a_0^2-\omega_v^2}\sin{(a_0t+\phi_0)}\sin{(\omega_vt+\delta)}\right.\\
 \left. +\frac{a_0a_1+
\omega_va_2}{a_0^2-\omega_v^2}\cos{(a_0t+\phi_0)}\cos{(\omega_vt+\delta)}\right]\xi_\|,
 \end{array}\label{eq10}\end{equation}
where $\xi_\bot=<\xi_z(t)>$ and $\xi_\|=\sqrt{1-\xi_\bot^2}$.
Angular brackets mean the time average. If $|a_0|$ is close to
$\omega_v$, the corrections to the spin dynamics given by Eq.
(\ref{eq10}) can be large. Therefore, the condition
$|a_0|=\omega_v$ defines a resonance. When the spin rotation
frequency is far from the resonance, Eqs.
(\ref{eq1p}),(\ref{eq2p}),(\ref{eq9}),(\ref{eq10}) lead to the
following expression for the average angular velocity of spin
rotation:
\begin{equation}
\begin{array}{c}
<\dot{\psi}>=\lambda\Omega_0(1-C_F), ~~~
C_F=\frac14\left[1-\frac{\Omega_0^2}{\gamma^2(\Omega_0^2-\omega_v^2)}-
\frac{\omega_v^2(f-1)(f-1+2/\gamma)}{\Omega_0^2-\omega_v^2}\right]\psi_0^2.\\
\end{array}\label{eq12}\end{equation}

This formula coincides with the result obtained in Refs.
\cite{FPL,FF}. However, in the real g$-$2 experiment
$\omega_v/\Omega_0\sim10$. Therefore, we may not average over the
spin rotation which is rather slow. As a result of averaging only
over the vertical betatron oscillation, Eq. (\ref{eq12}) takes the
form
\begin{equation}
\begin{array}{c}
<\dot{\psi}>=\lambda\Omega_0(1-C),   ~~~ C=C_F+\frac{(\gamma-1)^2
\Omega_0^2-f^2\gamma^2\omega_v^2}{4\gamma^2(\Omega_0^2-\omega_v^2)}
\cos{[2(\Omega_0t+\lambda\phi_0)]}\psi_0^2
\cdot\frac{1+\xi_\bot^2}{1-\xi_\bot^2}.
\end{array}\label{eq18}\end{equation}
Since the average value of the last term oscillating with the
angular frequency $2\Omega_0$ is zero, formula (\ref{eq18}) is in
the best agreement with the previously obtained result. However,
it is reasonable to include the oscillatory term in the fitting
process instead of eli\-mi\-na\-ting it. In the real g$-$2
experiment, $f\approx1,~\xi_\bot=0,$~$\gamma\gg 1$, ~$\phi_0$
equals $\pm\frac{\pi}{2}$, and the coefficient $C$ is given by
$$
C=\frac14\left[1-\cos{(2\Omega_0t)}\right]\psi_0^2.$$ The
quantities $C_F$ and $C$ are rather small because they are
proportional to $\psi_0^2$. In the real g$-$2 experiment, $\psi_0$
is less than $1\cdot10^{-3}$.

The vertical betatron oscillation violates the sinusoidality of
spin motion.

Essentially, the result of averaging over the vertical betatron
oscillation is independent of an initial phase of this
oscillation. Therefore, coherent and incoherent oscillations give
the same effect. The oscillation of the angular velocity
$<\dot{\psi}>$ averaged in such a way is the result of the fixed
initial spin direction.

The quantity $O$ in Eqs. (\ref{eq1p}) and (\ref{eq2p}) is much
smaller than the corresponding quantity $o$ in Eqs. (\ref{eq2})
and (\ref{eqo}). A significant difference between these quantities
is conditioned by a relative smallness of $N_\rho$ and
$\omega_\phi$ in comparison with $\xi_\rho$ and
$\left(\omega_a\right)_\phi$. Such a smallness is due to the fact
that the radial component of particle momentum is proportional to
the small quantity $\rho_0$ because it is caused by the radial
oscillation. The radial component of spin given
by Eq. (\ref{eqno}) is defined by the g$-$2 precession. 
The importance of $O$ has been first found by Granger and Ford
\cite{GF} and formula (\ref{eq12}) currently used has been derived
by Farley \cite{FPL}.

\section {DISCUSSION}

The particle EDM causes the spin rotation about the radial axis.
As a result, the spin precession axis becomes tilted. If the spin
rotation in the horizontal plane is cancelled or strongly
restricted, the spin acquires a vertical component and the angle
between the spin and the horizontal plane linearly increases from
zero with time \cite{Sem}. Therefore, the vertical magnetic field,
the radial electric one, and magnetic focusing are expected to
govern the particle motion in the EDM experiment. The best
restriction of spin motion in the horizontal plane is obtained on
condition that the radial electric field is adjusted to
$$ E=E_0=\frac{a\beta\gamma^2}{1-a\beta^2\gamma^2}B $$
(see Ref. \cite{EDM}). Such a restriction permits the
high-precision detection of the EDM.

Field misalignments and beam oscillations cause the spin motion
imitating the presence of the EDM. In particular, the electric
field may not be perfectly in the horizontal plane. In this case,
the appearance of a vertical electric field leads to the vertical
deflection of spin simulating the EDM effect \cite{EDM,EDMP}. The
use of Eq. (\ref{eq8}) simplifies the quantitative analysis of
such effects. In this equation, the last but one term describes
the spin rotation about the radial axis. The preceding terms also
affect this rotation. The second-order effects discussed in Ref.
\cite{EDM} can also be calculated by means of Eq. (\ref{eq8}).
This equation determines the general approximate relation between
the effects of AMM and EDM:
$$
\frac{|\bm\Omega_{EDM}|}{|\bm\omega_a|}\approx\frac{\beta|\eta|}{2|a|}
=\frac{\beta|d|}{|\mu'|}. $$

The EDM might cause a small tilt of the spin precession axis of
muons in the g$-$2 experiment \cite{FS,MN}. Measuring this tilt
sets a limit of $2.8\times 10^{-19}$ e$\cdot$cm on the muon EDM
\cite{MN}. If the muon EDM were equal to this limit, it would
stimulate the rotation about the radial axis with the angular
frequency $|\bm\Omega_{EDM}|=3\times10^{-3}|\bm\omega_a|$.

The first excellent example of analytical description of
complicated spin dynamics in storage rings is given in Refs.
\cite{GF,FPL,FF}. In these works, the effects of the vertical and
radial CBOs on the horizontal motion of spin in the g$-$2
experiment are described. For this purpose, the particle rest
frame was used. However, this frame is inconvenient for taking
into account the synchrotron motion and the particle acceleration
caused by a locally nonzero longitudinal electric field.
For example, such a particle motion is important in the EDM
experiment. If a strong radial electric field is used
\cite{EDM,EDMP}, its imperfection results in the appearance of a
longitudinal component leading to a systematical error. In Refs.
\cite{FPL,FF} and other works, the effect described by the term
$o\bm e_z$ in Eq. (\ref{eq7}) was ignored. The description of
spin motion given in these works is therefore approximate, while
Eq. (\ref{eq7}) is exact.

Obtained equations (\ref{eq5}),(\ref{eq7}), and (\ref{eq8})
describing the spin dynamics in cylindrical coordinates have been
derived in the general form. They can be helpful for
high-precision experiments at storage rings when the spin motion
needs to be analytically calculated. Nevertheless, these equations
are convenient only when the ring is either circular or divided
into circular sectors by empty spaces. Moreover, the analytical
description given by these equations can be useless if the
configuration of the main fields is not simple enough. In this
case, it is necessary to use computer calculations. The algorithms
of such calculations can handle all kinds of common magnet types
and also the effects of misalignments. These algorithms are based
on different coordinate systems (e.g., on Frenet-Serret
coordinates \cite{CR}). The method for calculating the spin motion
in storage rings was developed by Derbenev, Kondratenko, and
Skrinsky \cite{DKS} (see also Ref. \cite{Sh} and references
therein). Another method using the Frenet-Serret curvilinear
coordinates was proposed by Courant and Ruth \cite{CR}. These
methods have been successfully used in various experiments and are
also applicable in any other case.

However, the analytical description of spin dynamics can be
necessary for several high-precision experiments. Only the
analytical description can guarantee the clear understanding
needed for the g$-$2 and EDM experiments. The use of Eqs.
(\ref{eq5}),(\ref{eq7}), and (\ref{eq8}) for analytical
calculations of spin dynamics in these experiments can be very
successful. Other formalisms seem to be less convenient. It is
natural to use the cylindrical coordinate system for analyzing the
spin motion in circular storage rings. Therefore, the obtained
exact equation of spin motion in cylindrical coordinates can also
be utilized in other cases where the configuration of the main
fields is simple enough.

\section {SUMMARY}

The exact formula for the angular velocity of particle motion in
the horizontal plane has been obtained. The exact equation
defining the spin motion in the cylindrical coordinate system with
allowance for the particle EDM has been derived. This equation is
convenient for analytical calculations of spin dynamics when the
configuration of the main fields is simple enough. Such
calculations can be needed for several high-precision experiments.
Cylindrical coordinates can be used if the ring is either circular
or divided into circular sectors. Averaging of the angular
velocity of spin rotation has been performed. The generalized
formula for the influence of the vertical betatron oscillation on
the angular velocity of spin rotation in the g$-$2 experiment
(pitch correction) has been found. This formula agrees with the
previous result \cite{FPL,FF} and contains the additional
oscillatory term which can be used for fitting. The relative
importance of the terms in the equation of spin motion including
the EDM-dependent ones is discussed.

\section* {Acknowledgements}

The author would like to thank F.J.M. Farley, Y.F. Orlov, Y.K.
Semertzidis, and Yu.M. Shatunov for helpful remarks and
discussions. The author also wishes to thank the Brookhaven
National Laboratory for the invitation and support.

\end{document}